# Unexpected magnetism explained in Cu/Cu2O-rGO nanocomposite

Rajarshi Roy[a, †,1], Kaustav Bhattacharjee[a,‡,1], Satya Prakash Pati[b], Korak Biswas[c], Kalyan Kumar Chattopadhyay[a]

[a] Thin Film and Nanoscience Laboratory, Department of Physics, Jadavpur University, Kolkata- 700032, India.

[b] Department of Electronic Engineering, Graduate School of Electronic Engineering, Tohoku University, Sendai 980-8579, Japan.

[c] Department. of Physics, Indian Institute of Science Education and Research, Pune, 411008, India.

[‡] Centre for Materials for Electronics Technology (C-MET), Ministry of Electronics and Information Technology (MeitY), Panchavati, Off Pashan Road, Pune 411008, India.

[†]Present address: Institute of Physics of the Czech Academy of Sciences, 10/112 Cukrvarnická, Prague 162 00.

[1]These authors contributed equally to this work.

## Abstract

The observation of room temperature ferromagnetism along with a low temperature paramagnetic counterpart in undoped Cu-$Cu_2$O-rGO nanocomposite was demonstrated. A phenomenological approach was taken to explain the observations based on 3D Ising model for arbitrary spins generated due to Cu vacancy in the $Cu_2$O system preferably at the interface.

**Introduction**

In semiconductor devices, where the charge of the electron usually plays a key role, further maneuverability of spin direction might open up fresh avenues pertaining to compound architectures; and consequently the development of nanocomposite materials with both charge and magnetic moments have incrementally established themselves as the foci of the current spintronic paradigm.[1] In comparison to the magnetism in undoped n-type semiconductor (such as, ZnO, $TiO_2$)[2,3] the p-type cuprite ($Cu_2O$) material has been less explored. Observation of room temperature ferromagnetism (RTFM) in $Cu_2O$ would not only help to fabricate spintronic material for future applications but also offer a fundamental platform to study its microscopic origin. Pure $Cu_2O$ is a diamagnetic material where neither $Cu^{+1}$ nor $O^{2-}$ ions are magnetic. However, RTFM in undoped $Cu_2O$ fine powder,[4,5] nanowires[6] and $CuO/Cu_2O$[7] or $Cu/Cu_2O$ interface[8] has recently been reported in the literature. Theoretically, the ferromagnetism in all these materials is claimed to be induced by the oxygen interstitial and/or Cu vacancy, formed during the annealing or during the preparation under controlled oxygen partial pressure condition. In fact, $Cu_2O$ was among the first reported solids for which a dependence of the optical and electrical property on the preparation condition was observed. The deviation from stoichiometry, δ (in $Cu_{2-δ}O$) is strongly dependent on oxygen activity.[9]

From first principles calculations, Soon *et al*[10] has confirmed the source of FM ordering in sub-stoichiometric $Cu_2O$ arises mainly due to cation deficiency. Specifically for oxygen lean case a low energy interfacial structure, $Cu_2O$ (111)-Cu, is formed, which contains Cu vacancy (coordinatively unsaturated) at the surface. Recently, Li *et al*[11] has further exemplified the impact of interfacial Cu vacancy for ferromagnetic modulation in Cu(111)/$Cu_2O$(111) interface.

In this letter, we report RTFM in Cu/$Cu_2O$-rGO nanocomposite owing to its intrinsic defects (Cu vacancy in the $Cu_2O$ lattice) at the Cu/$Cu_2O$ interface and explained its microscopic origin based on the Ising model for arbitrary spins interacting with all nearest neighbor interactions (NN) in a 3D square lattice. Simulation not only shows good approximation of the experimental data at high temperature, but also, explains the unusual low temperature paramagnetic contribution coming from the uncorrelated (lonely) spins. Further, a 3D representation of defective lattice was demonstrated from the knowledge of point defects in the $Cu_2O$ lattice.

**Methods**

The sample was synthesized using the following protocol. In brief, 80 mL of rGO (1 mg/mL) dispersion in hydrazine and ammonia (1:7 vol %) water was prepared by reported procedure.[12] Then 0.17 g (1 mM) of $CuCl_2$ was added to this solution and stirred until properly mixed. Following this, the whole solution was added in a Teflon lined autoclave and heated at 180 °C for 24 h. The existing presence of hydrazine hydrate in the RGO solution helps to reduce the $Cu^{+2}$ ions and $Cu/Cu_2O$-rGO nanocomposite was formed in the in situ hydrothermal reaction. The resultant powder was collected, washed with water and dried in a vacuum oven for further characterization. It is to be noted that the presence of rGO helps to obtain a better control over the size distribution of particles by trapping the guest nanoparticles due to its large contact area.[13]

Sample was characterized for phase identification by high resolution powder X-ray diffraction (PXRD) (Rigaku Ultima-III) method using Cu kα (λ = 0.154 nm) radiation in θ/2θ mode. X-ray photoelectron spectroscopy (XPS) analysis was done using monochromatic Al Kα (hv = 1486.6 eV) X-ray source with a hemispherical analyzer by SPECS, HSA 3500. Morphological and microstructural characterizations are conducted using Field Emission Scanning Electron Microscopy (FESEM) (Hitachi S-4800) and High-Resolution transmission Electron Microscopy (HRTEM) (JEOL JEM – 2100) operating at 5 kV and 200 kV respectively. Magnetic characterizations were done with Superconducting Quantum Interference Device (SQUID) from Quantum Design.

**Result and discussions**

Figure 1(a) depicts the XRD pattern of the nanocomposite samples showing a quantitative phase analysis carried out by MAUD.[14] Two phases, viz., cubic Cu (SG: Fm-3m) and cubic $Cu_2O$ (SG: Pn-3m:1) has been identified with their relative abundance of 81 % and 19 % respectively. It is reported that employing RGO as an electron acceptor can enhance the stability of $Cu_2O$ in a composite.[15] It has been theoretically shown that the subsurface $Cu_2O$ could be an stable intermediate during the reduction of CuO, and $Cu^{2+}$ preferably formed $Cu^+$ before it is directly reduce to metallic $Cu^0$.[16] Thus in the present case, the role of rGO was also to stabilize the (+1) state of Cu during the reduction process. For a face centered cubic structure (like Cu or $Cu_2O$) the thermodynamic approach to minimize the interfacial free energy (γ) of the lattice planes follows the order, γ{111} <γ{100} <γ{110}.[17] However, Cu and $Cu_2O$ differ significantly in crystal structure by the presence of oxygen atom in the interstitial tetrahedral sites for $Cu_2O$ which leads to the almost 18 % lattice expansion. As defects induce magnetism is well reported in an undoped oxide semiconductors, we calculate the type of defects present in the $Cu_2O$ lattice from Rietveld X-ray crystal structure refinement analysis. In the present case, refinement of the Cu occupancy factors for the $Cu_2O$ phase shows a significant amount (9 %) of deficiency ($Cu_{(2-0.18)}O$) in the sample. Since, under present reaction condition Cu has to distribute between two crystallographic phases, such cation deficiency is expected to occur at the cost of lattice expansion at the $Cu/Cu_2O$ interface. Additionally, planar defects can also be incorporated into the $Cu_2O$ lattice owing to any deviation from the parallel epitaxial relation between the Cu and $Cu_2O$ fcc crystal structure. Lutterotti and Gialanella introduced a novel method to quantify such type of defects and implemented this in the Rietveld's refinement software MAUD.[18] These faults were expressed in terms of probability, which contributes to the peak shift, anisotropic broadening and asymmetry. In the present case, the $Cu_2O$ (111) peak could be better fitted both in terms of intensity and FWHM through introduction of additional anisotropic planar defect (PD) term in the refinement procedure according to the method described by Warren[19] (see figure S1, supporting information). Analysis shows significant amount of deformation fault probability ($\alpha = 0.0018$), growth probability ($\alpha'' = 0.0018$) and twin fault probability ($\beta = 0.0029$) due to the off stoichiometry in the sample. All the refined parameters can be found in table S1 in supporting information. In order to substantiate the chemical composition of the sample, we carried out high resolution XPS analysis at the Cu 2p and O 1s spectral region as shown in figure 2(b). The representative survey scan was shown in figure S2, supporting information. The spectrum was calibrated against C 1s peak positioned at 284.6 eV.[20] The peak profile of Cu $2p_{3/2}$ spectrum was found to be asymmetric with major peak positioned at binding energies of 932.4 eV. Though there is considerable variation in the literature of the binding energy for $Cu_2O$, nevertheless, this can be assigned for the $Cu^+$ or $Cu^0$ (as both has almost overlapping binding energy in this region).[20,21] An additional peak at binding energy of 933.3 eV which appear as a shoulder signifies contribution from Cu vacancy in the sample.[20,22] Nevertheless, the symmetric nature of the O 1s spectrum peaking at 530.6 eV, attributed to $O^{2-}$ in $Cu_2O$ lattice.[21]

Figure 2(a) and 2(b) depict the morphological characterizations of the sample using FESEM and TEM, respectively. The composite character was revealed in surface analysis of the sample; nanoparticles with average particle size roughly ~50-80 nm distributed over the rGO sheets. Similar adhesion of particle/particle aggregates on the RGO sheet were also found in TEM analysis. We have previously discussed the role of rGO as a matrix for obtaining well distributed nanoparticles. Figure 2(c) shows the HRTEM inspection on the surface of the rGO sheet which evokes the fine information of the nanocomposite. The linear dark and light bands in the HRTEM image indicate twin faults within the metallic nanoparticles (Cu, in the present case) core. From the calculation of inter planer distances on different regions in the figure 2(c) (a magnified portion of the selected area was shown in the figure 2(d)) it reveals that, the more dense region in the TEM image are made of metallic Cu(111) crystal habit while formation of $Cu_2O$ was noticed mainly on the particle surface or particle-particle interface region. The line profile analysis (see figure S3, supporting information for detail) at the interface further uncover that 7 (111) planes of Cu matches well to 6 (111) plane of $Cu_2O$, indicating that the interface has a 7x6 lattice matching relationship.[23] However, because of the large misfit between the lattice parameters of Cu and $Cu_2O$, a good matching of the two crystal structure at the interface can be achieved by introducing defects that would allow to deviates from the ideal parallel epitaxial relationship between the two planes.

The field dependence of magnetization, M(*H*) of the composite sample measured at different temperature was shown in the Fig. 3(a). The data were presented after substarcting the diamagnetic contribution from both Cu metal (taken from literature) as well as from rGO (calculated from the experimental M-H curve at room temperature, see figure S4 in the supporting information) from the sample holder. The sample appeared to be ferromagnetic (with non-zero coercivity and remanent magnetization) at room temperature. RTFM in undoped nonmagnetic oxides has invoke a matter of great controversy in material science since its discovery.[24] As both the Cu and $Cu_2O$ are diamagnetic in nature, in the present case, we consider the magnetism is coming from the defects in $Cu_2O$ lattice (mainly Cu vacancy) incorporated during the sysnthesis procedure. Interestingly, from figure 3(a) we can see an anomalous paramagnetic susceptibility contributing to the high field region of the M-H curve, below 100 K. At the same time, both remanent magnetization ($M_r$) and coercivity ($H_c$), as obtained from the low field region of the each isothermal M(H) curves, exhibit a rapid increase (change in slope) below 50 K

(as shown in the inset of figure 3(a)). Hence, an enhanced ferromagnetic property is indicated all together. The temperature dependence of magnetization, M(T), measured at temperature interval 300-3 K under a constant applied field of 100 Oe was shown in figure 3(b). The field cooled (FC) and zero field cooled (ZFC) curves were clearly separated until the temperature above 300 K, which suport the existance of RTFM.[25] It can be noted that no blocking temperature occur in this temperature range, indicating that there is no superparamagnetic cluster in the sample.[25] FC-ZFC measurements performed at low temperature range 100-3 K and higher applied magnetic fields (See inset of Fig. 3(b)) show a general increase in magnetization with decrease in temperature which is quite obvious for paramagnetic species. This observation agrees with the previous results reported by O'Keeffe and Stone for copper-oxygen system, where paramagnetism was reported in cuprite material prepared by low temperature (below 160 °C) and low oxygen pertial pressure condition and explained on the basis of vacancies or defects on the surface.[26]

To understand the microscopic origin of such system, we consider the Ising model which in its simple form even, is still of current interest.[27] The model we consider in this work, similar to the model consists of points (or spins, +1 or -1) arranged in a regular cube lattice,[28] but differ in the position of spin variables; these spins (as defined as defect spins) were introduced by only a fraction of the total lattice sites at any point in the cube lattice considering all nearest-neighbour (NN) interactions. The popular choice being start from a random configuration, the system will have to evolve for several Monte Carlo (MC) steps before an equilibrium state at a given temperature is obtained. We consider the periodic boundary condition and the Hamiltonian of the system is given by,

$$H = -J_{ij} \sum_{<ij>} s_i s_j - h \sum_i s_i$$

where, $J_{ij}$ is the coupling parameter between the adjacent atoms, h is external field strength and $s_i$ is the spin of particle. The initial configuration was generated by placing spins (both spin up and spin down) randomly on a cube (L=$150^3$) lattice and performed $10^6$ MC steps (spin flip-flop move) to obtain the minimum energy of the system at particular temperature. The move is based on energy considerations and is done by Metropolis algorithm. All quantities were calculated in normalised unit. $k_B$ is taken 1 unit and J is taken 1 unit. The whole procedure was done for different temperature ranging from 0.1 to 20. The simulation code was written in Python script. In this case we consider defect spin fraction of 9 % under weak external applied magnetic field (not strong enough to flip-flop the spin) to study the spin organization after equilibrium has reached. Magnetization per lattice site was calculated in the presence of small external magnetic field. We then verify the behaviour of the magnetization with temperature as shown in figure 3(c). The magnetization behaviour indicates a paramagnetic response at low temperature, similar to that in our experiment, suggesting the spins are begin to freezes at a particular direction at this temperature region giving rise to high magnetization value. To understand the reason behind such behaviour, we calculate the loneliness factor (P) for each spins, which is defined as, P = 1-(total NN/6), the maximum being P =1 (no NN) and minimum is P=0 (6 NN).We observed that, the spins which have higher loneliness are freezes more rapidly at a particular direction with decrease in temperature. These isolated spins are actually responsible for the low temperature paramagnetic behaviour in the system. Figure 3(d) demonstrate a 2D representative lattice of the 3D Ising model generated after the MC simulation at two different temperatures where the defect spins are coupled by all nearest neighbour interactions. One can clearly visualize that a particular type of spin configuration is predominate over the other at low temperature and it is the more lonely spins which freezes sharply to a perticular direction, thus leading to higher magnetization values at high field.

**Conclusions**

In summary, we reported the unexpected observation of RTFM in otherwise non-magnetic undoped Cu/Cu$_2$O-rGO nanocomposite material. Crystal structural and microstructural analyses show significant amount of Cu vacancy within the Cu$_2$O lattice preferably situated at the particle surface or particle-particle interface. More importantly, a low temperature paramagnetic response was noted in the present case which can be well explained on the basis of 3D Ising model for randomly distributed spins associated with poor nearest neighbour interaction.

**Conflicts of interest**

There are no conflicts to declare.

**Acknowledgement**

One of the authors (RR) would like to acknowledge the Council of Scientific and Industrial Research (CSIR), Govt. of India, for awarding him a Senior Research Fellowship (SRF) during the tenure of this research work. The authors are also grateful to the UGC (for UPEII programme) and the DST, the Govt. of India for financial support.

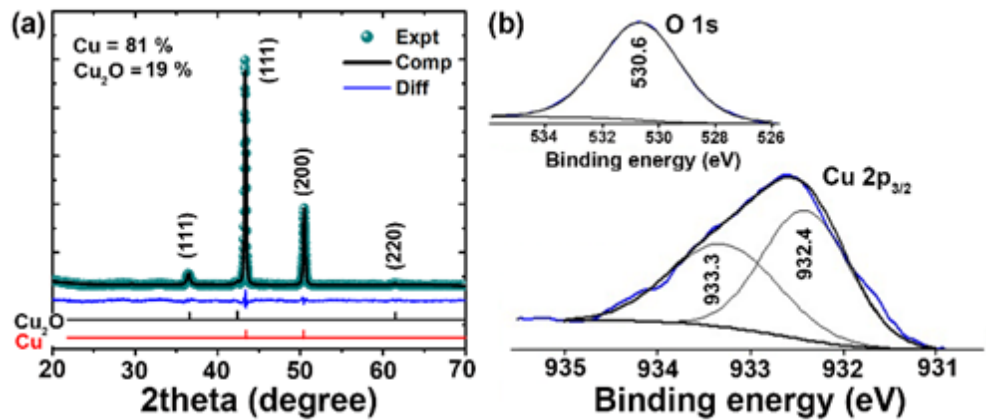

**Fig. 1** (a) PXRD pattern of the Cu/Cu$_2$O-rGO nanocomposite showing the Rietveld refinement fit using MAUD considering the PD term and other associated free parameters. Details of the fitting parameters can be found in the supporting information. The R$_{wp}$, R$_{exp}$, *GoF* were given by 1.24, 0.85 and 1.46 respectively. (b) High resolution XPS scan for Cu 2p$_{3/2}$ and O 1s spectral region.

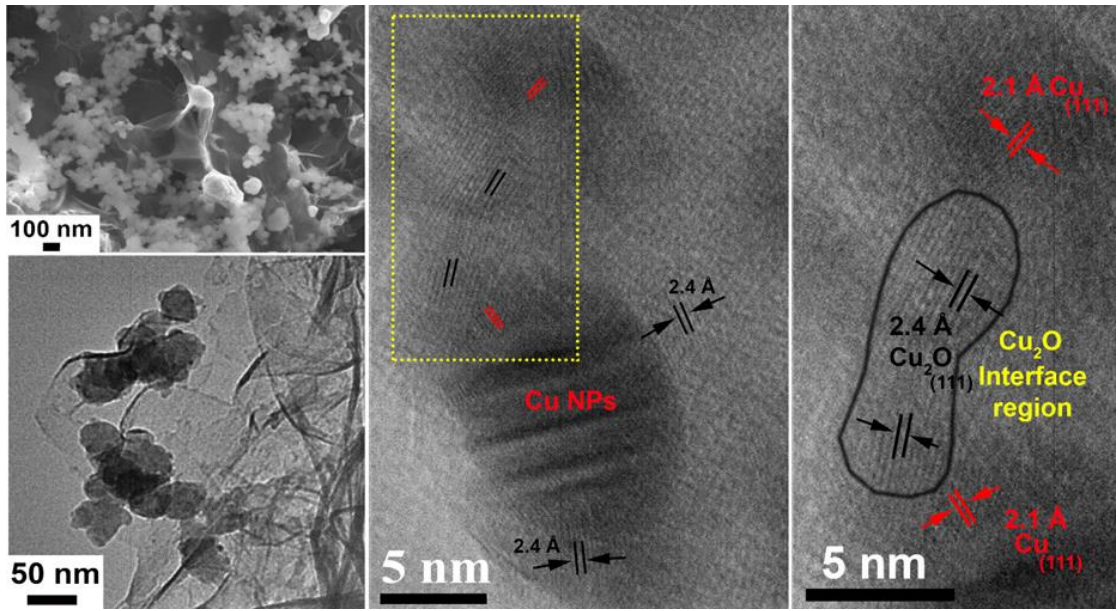

**Fig. 2** (a) FESEM image and (b) TEM image of the Cu/Cu$_2$O-rGO nanocomposite material. (c) HRTEM image taken at the edge of the RGO sheet. The red (2.1 Å) and black (2.4 Å) line represent lattice spacing for (111) plane of Cu and Cu$_2$O, respectively. (d) A zoom image of the selected region (rectangle in figure (c)) showing the particle-particle interface region is dominated by Cu$_2$O.

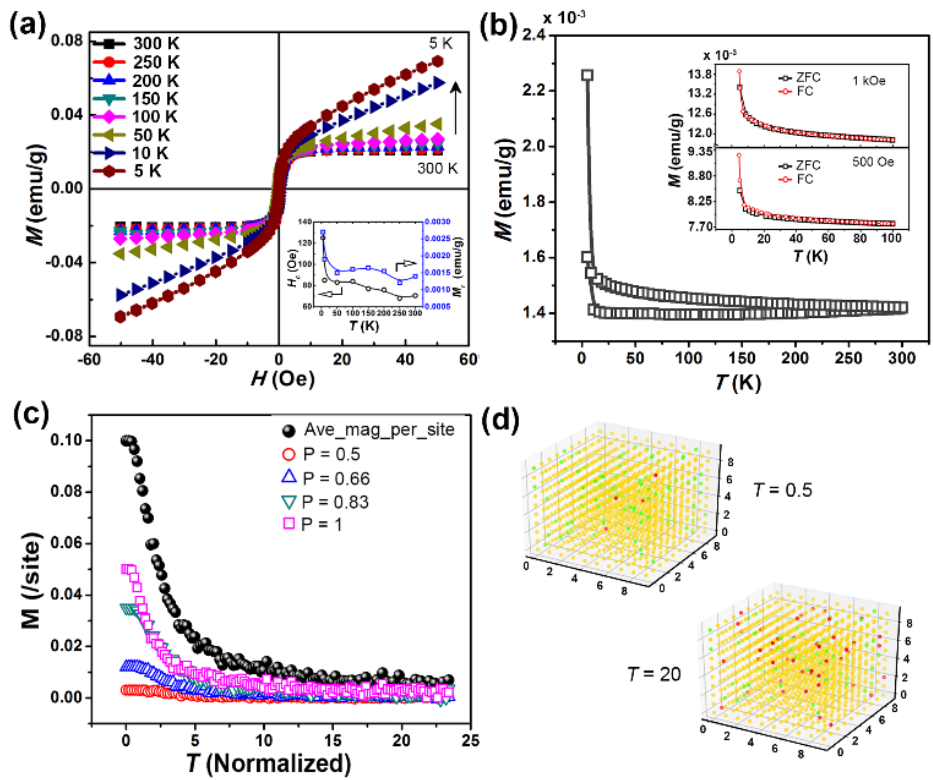

**Fig. 3** (a) M(H) curves for the Cu/Cu$_2$O-rGO nanocomposite recorded at different temperatures. The variation of H$_c$ and M$_r$ with measuring temperature were shown in the inset. (b) M(T) curve for the Cu/Cu$_2$O-rGO nanocomposite at an external magnetic field of 100 Oe. Inset show the FC-ZFC curves taken at external magnetic field of 500 Oe and 1 kOe. (c) Simulated average magnetization per site vs. temperature plot; comparison of the effect of temperature on magnetization due to different lonliness (P) spin values. (d) 3D representative images of the (considering 9 % defect concentration) Issing lattice geberated after complete thermal equlibrium at two different temperatures. Yellow, red and green spots represents the lattice points, spin up (+1) and spin down (-1) respectively.

# Supplementary Information (ESI)

## Unexpected magnetism explained in Cu/Cu$_2$O-rGO nanocomposite

Rajarshi Roy[1], Kaustav Bhattacharjee[1], Satya Prakash Pati, Korak Biswas, Kalyan Kumar Chattopadhyay

**Powder XRD refinement analysis:**

The PXRD refinement program (Material Analysis Using Diffraction, MAUD 2.26, by Luca Lutterotti) we used in this study for crystal structure analysis is based on the previous literature.[1-5] The background was fitted with a polynomial function of degree 4. The peak shape was assumed to be pseudo-voigt (cagloiti pV) function. The Marquardt least squares procedures were adopted for the minimization of the difference between the observed and simulated powder diffraction patterns and the minimization was monitored using the reliability index parameters, $R_{wp}$ (weighted residual error), and $R_{exp}$ (expected error). Refinement continues till convergence is reached with the value of the quality factor, goodness of fit (*GoF*) very close to1 (varies between 1.1 and 1.7).[6]

$$R_{wp} = \left[ \frac{\sum_i w_i(I_o - I_c)^2}{\sum_i w_i I_o^2} \right]^{1/2}$$

$$R_{exp} = \left[ \frac{(N-P)}{\sum_i w_i I_o^2} \right]^{1/2}$$

where, $I_0$ and $I_c$ are the experimental and calculated intensities respectively, $w_i$ (= $I/I_0$) and N are the weight and number of experimental observations and P the number of fitting parameters. This leads to the value of *GoF*$^3$:

$GoF = R_{wp} / R_{exp}$

The weight fractions of the component phases were obtained from the following equation[7]

$$W_i = \frac{S_i \rho_i V_i^2}{\sum_j S_j \rho_j V_i^2}$$

where, $W_i$ is the weight fraction, $S_i$ is the scale factor, $V_i$ is the unit cell volume, $\rho_i$ is the density of phase *i*, and subscript *j* includes all phases present

**Table S1:** Values of Rietveld refinement parameters for the Cu/Cu$_2$O-rGO sample. Values in the parenthesis are estimated standard deviation.

| Phase | Copper (Cu) (Cif file No. 4105040) | | | Cuprite (Cu$_2$O) (Cif file No. 1010941) | | |
|---|---|---|---|---|---|---|
| Volume fraction (%) | 81 | | | 19 | | |
| Crystal system | Cubic | | | Cubic | | |
| Space group | Fm-3m | | | Pn-3m:1 | | |
| Lattice parameter (Å) | 3.6158 (3) | | | 4.2664 (7) | | |
| dCu-O (Å) | | | | 1.84 | | |
| dCu-Cu (Å) | 2.55 | | | 3.01 | | |
| dO-O (Å) | | | | 3.69 | | |
| Cu-O-Cu (°) | | | | 109.4 | | |
| Cu occupancy factor | 1 | | | 0.91 (6) | | |
| Fractional coordinate of Cu | x | y | z | x | y | z |
| | 0 | 0 | 0 | 0.25 | 0.25 | 0.25 |
| Planer defects | | | | α′ = 0.0018 | α″ = 0.0018 | β = 0.0029 |

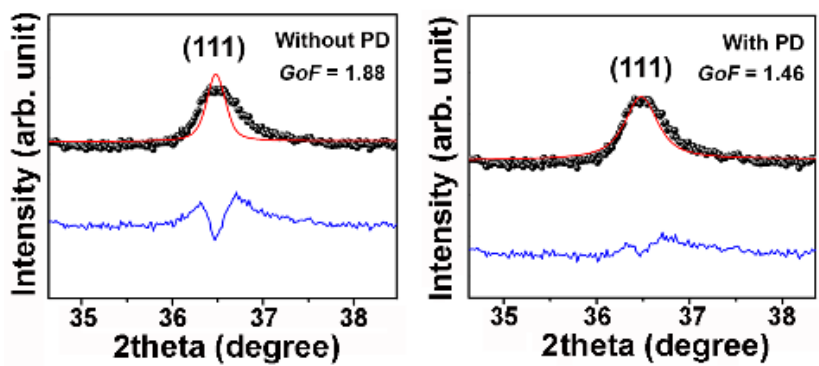

**Figure S1:** Fitting of Cu$_2$O (111) peak with and without planar defect (PD). The change in fitting quality was clearly seen by eyes as well as from *GoF* values.

**XPS analysis for the sample.**

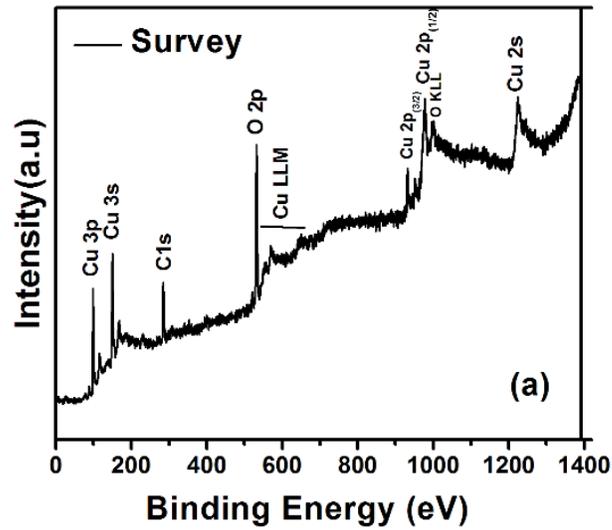

**Figure S2:** XPS survey scan for the $Cu/Cu_2O$-rGO nanocomposite.

**TEM analysis of the sample.**

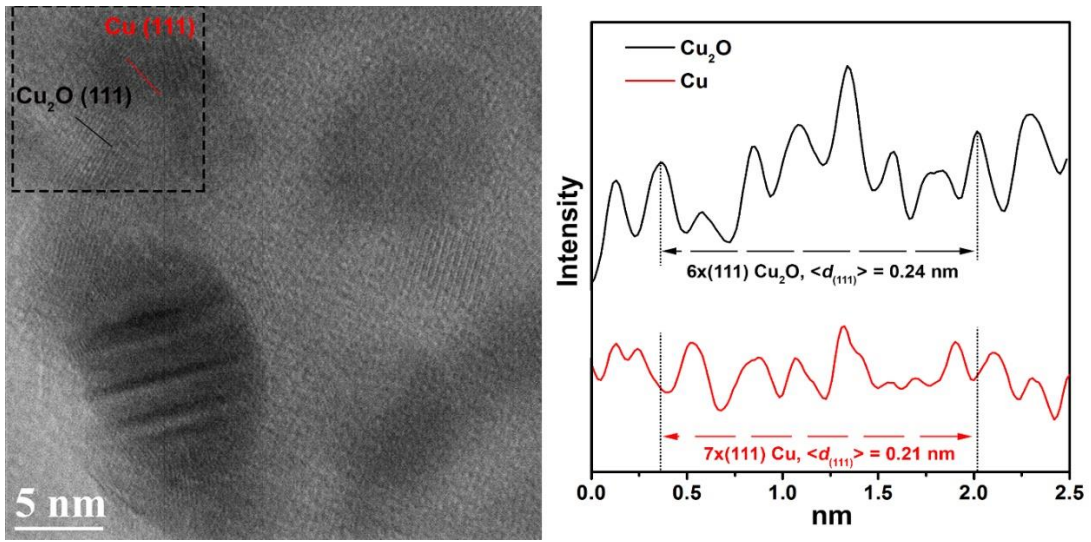

**Figure S3:** Line profile analyses from the red and blue line in the TEM image (left panels) with annotations indicating the lattice matching of the $Cu(111)/Cu_2O(111)$, with a 7×6 relationship.

**M-H analysis for rGO powder**

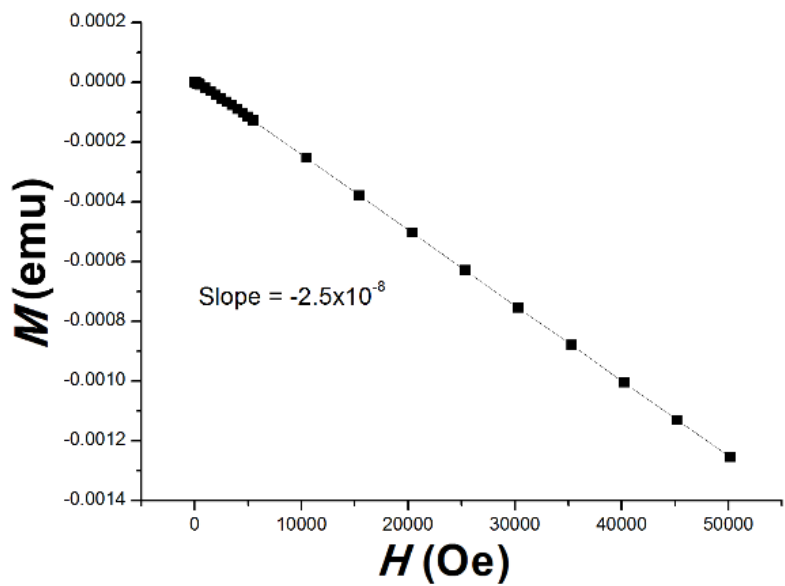

**Figure S4:** Room temperature M-H curve for pure rGO